# Decoupling Trap States in Amorphous Oxide Semiconductor Transistor Using a Conductive–Semiconducting–Conductive MOS Structure

Kai Chen[1,2,3], Wei Zhang[2], Xuanyao Fong[2], Shengman Li[2*] and Yunlong Li[1,3*]
[1]College of Integrated Circuits, Zhejiang University, Hangzhou 311200, China. [2]Department of ECE, National University of Singapore, Singapore 110573. [3]Zhejiang ICsprout Semiconductor Co., Ltd, Hangzhou 311200, China.
[*]Corresponding authors; Email: shengman_li@nus.edu.sg, bta@zju.edu.cn

***Abstract*—**We propose a novel amorphous oxide semiconductor (AOS)-based conductive–semiconducting–conductive (CSC) MOS test structure to effectively sense, discriminate, and analyze distinct trap-induced conductance losses in AOS transistors. In conventional test structures, non-ideal metal-AOS contacts convolute the conductance loss generated by various trap states. The CSC architecture mitigates these contact-related artifacts, enabling the decoupling of interface states and shallow states. Furthermore, the contributions from leakage, series resistance, and device-geometry effects are identified and calibrated. Our results yield bias-resolved trap profiles and reveal three distinct regimes: (1) In the depletion region, interface states dominate the conductance loss. These states exhibit sub-microsecond time constants and their density increases slightly with deeper depletion, which is associated with donor-like states. (2) From weak to moderate accumulation, shallow states become the primary contributor, displaying a Gaussian-like volumetric density-of-states distribution with response times spanning from microseconds to nanoseconds. (3) In the strong accumulation regime, interface states regain dominance with an increased state quantity, while their time constants remain invariant. These bias-resolved profiles clarify how different trap populations impact AOS transistors under distinct operating conditions, providing intuitive metrics for understanding reliability degradation. Consequently, the CSC structure offers a simple, scalable, and robust platform for trap characterization in AOS transistors.

## I. INTRODUCTION

Amorphous oxide semiconductors (AOS) combine low-temperature processability [1], ultralow off-state leakage [2], and competitive mobility [1,3], making them promising channel materials for monolithic 3D integration [4-6]. However, device reliability remains a major limitation, largely because of charge trapping in the AOS bulk and at the gate-oxide/AOS interface [7, 8]. As illustrated in **Fig. 1(a)**, AOS contains band-tail, shallow, deep, and near-valence-band states, while the gate stack additionally introduces interface traps [9]. Although bulk trap distributions have been extensively investigated through complementary physical and electrical characterization [10, 11], the electrical responses of bulk and interface traps are generally convoluted because both are coupled to the same channel. This ambiguity hinders identification of the dominant degradation mechanisms and the rational optimization of AOS transistor reliability.

The conductance method is widely used to characterize interface traps in Si MOS structures [12]. By applying a small AC signal over a range of frequencies, the conductance loss arising from the capture and emission of carriers by trap states can be measured. Combined with an appropriate equivalent-circuit model, the trap density and characteristic response time can be extracted [13]. Its application to nanometer-thick AOS channels, however, is strongly affected by the test structure. In conventional parallel-plate AOS MOS capacitors, nonideal metal–AOS contacts can introduce bias-dependent contact barriers and resistance, Fermi-level pinning, metal-induced gap states, and local degradation of the channel properties, thereby obscuring trap-related conductance loss [14], as shown in **Fig. 1(b)**. Edge-contact MOS capacitors have been proposed to alleviate these effects, but contact- and geometry-related limitations remain [15].

In this work, we propose a novel conductive–semiconducting–conductive (CSC) AOS MOS capacitor, as shown in **Fig. 1(c)**. By sufficiently oxygen-compensating the exposed channel between the source and drain while limiting oxygen compensation in the AOS underneath the source/drain electrodes, the contact regions retain a highly conductive state. The highly conductive contact regions enhance sensitivity to trap-induced conductance loss. Using this structure, the bias-dependent trap responses in the AOS MOS system are obtained and decoupled, as illustrated in **Fig. 1(d)**. Interface-related states dominate under depletion and strong accumulation, whereas shallow bulk states become prominent from weak to moderate accumulation. Specifically, under strong accumulation, the interface state density increases while maintaining sub-microsecond time constants, whereas the shallow states exhibit a Gaussian distribution with response times spanning from microseconds to nanoseconds. These results deepen the understanding of trap states in AOS transistors.

## II. FABRICATION OF CSC-AOS MOS CAPACITOR

**Fig. 2(a)** shows the cross-sectional and top-view schematics of the bottom-gate IGZTO transistor used to demonstrate the proposed CSC-AOS MOS capacitor, together with its detailed geometrical layout. The structure features a 5 μm overlap margin between the patterned IGZTO and source/drain regions on all sides. The fabrication process is illustrated in **Fig. 2(b)**. A 30-nm-thick Ni bottom gate was sputtered and patterned on a heavily doped p-type Si substrate with 100 nm $SiO_2$. A 6-nm-thick $HfO_2$ gate dielectric was then deposited by thermal ALD and annealed in $N_2$ at 400 °C for 30 min. Subsequently, a 4.5-nm-thick IGZTO channel was sputtered in Ar at room temperature. Ni/W/Pt source/drain electrodes were formed by sputtering and lift-off, where Ni provided low-resistance contact, W suppressed Ni–Pt interdiffusion, and Pt acted as an oxygen-blocking layer. Finally, $O_2$ annealing at 400 °C for 1 h oxygen-compensated the exposed IGZTO channel, while the Pt-covered IGZTO remained highly conductive, thereby forming the lateral conductive–semiconducting–conductive structure.

## III. RESULTS AND DISCUSSION

### *A. Electrical Confirmation of CSC Structure Formation*

The transfer characteristics of the CSC-IGZTO transistor are shown in **Fig. 3(a)**, exhibiting a steep subthreshold swing of approximately 63.6 mV/dec and a small hysteresis of about 4.7 mV. The multi-frequency C–V characteristics measured with the source and drain shorted are presented in **Fig. 3(b)**. Under negative gate bias, the exposed IGZTO channel between the source and drain is depleted, and the measured capacitance is therefore mainly attributed to the gate-dielectric capacitance coupled to the highly conductive IGZTO regions underneath the source/drain electrodes. The corresponding C–f characteristics in **Fig. 3(c)** remain nearly frequency independent throughout the depletion regime. When the drain terminal is left floating, as shown in **Fig. 3(d)**, the depletion-regime capacitance also remains nearly constant with frequency and is approximately half of that measured with the source and drain shorted, indicating that only one conductive overlap region is effectively sensed. These capacitance characteristics provide electrical evidence for the formation of the lateral CSC structure. The nominal exposed channel area and total source/drain overlap area are 400 and 900 $\mu m^2$ in this device, respectively. Based on the measured capacitance ratios and assuming a uniform areal capacitance, the effective semiconducting channel area is estimated to be approximately 600 $\mu m^2$. The larger-than-designed value suggests that part of the IGZTO beneath the source/drain edges was also oxygen compensated during the final annealing step. This electrically extracted effective area is therefore used in the subsequent trap-density calculations.

### *B. Effects of Resistance and Geometry on Trap Extraction*

To evaluate the influence of resistance-related effects on the conductance response, $G_P/\omega$–f measurements were first performed with the drain terminal floating, as shown in **Fig. 4(a)**. As the device enters accumulation, a pronounced peak emerges, increases with $V_G$, and shifts toward higher frequencies. The low-frequency $G_P/\omega$ component at 1 kHz initially increases with $V_G$ and subsequently decreases with frequency, consistent with the contribution of gate-dielectric leakage [13]. Its bias dependence closely follows that of the measured gate leakage, as shown in **Fig. 4(b)**. For an RC-related response, $f_{peak}\approx 1/(2\pi R_{eff}C_{eff})$. Because $C_{eff}$ varies only slightly over the investigated bias range, the peak position is mainly governed by $R_{eff}$. **Fig. 4(c)** shows a linear correlation between $f_{peak}$ and the total conductance ($1/R_{total}$) extracted from the transfer characteristics. Correspondingly, the extracted time constant follows the variation in $R_{total}$, as shown in **Fig. 4(d)**, supporting a predominantly RC-related origin of this peak. A weak high-frequency response is also observed near device turn-on in **Fig. 4(a)** but becomes progressively obscured by the stronger RC contribution at higher $V_G$.

The geometry dependence was further examined using CSC-IGZTO transistors with different channel lengths and shorted source/drain terminals. As shown in **Fig. 5**, increasing the channel length increases the channel resistance and leads to the emergence of additional $G_p/\omega$ peaks, demonstrating that distributed channel resistance can complicate trap extraction [15]. Meanwhile, the trap-induced conductance loss scales with the active channel area. For devices with insufficient area, the intrinsic trap response may therefore be masked by RC-related loss, gate leakage, and measurement noise. Subsequent trap-state analyses were performed on devices with sufficiently large effective channel areas.

### *C. Qualitative and Quantitative Analysis of Trap States*

Optical absorption and ultraviolet photoelectron spectroscopy (UPS) were used to establish the energy levels of the fabricated IGZTO film, as shown in **Fig. 6**. The extracted optical bandgap is 3.839 eV, while $E_C-E_F$ is approximately 0.283 eV. **Fig. 7(a)** and **7(b)** present the $G_p/\omega$ responses of the CSC-IGZTO transistor with L=5 μm, W=80 μm, and shorted source/drain terminals. The pronounced bias- and frequency-dependent spectra indicate contributions from trap responses with distinct characteristic time constants.

The low-frequency $G_p/\omega$ component increases with $V_G$ and is mainly associated with gate-dielectric leakage. As estimated from its 1 kHz value in **Fig. 8(a)**, this contribution decreases approximately as 1/f and becomes negligible above 100 kHz. In the depletion regime from $V_G$=−1 to −0.1 V, a nearly bias-independent high-frequency peak is observed in **Fig. 8(b)**, suggesting an interface-related origin. As $V_G$ increases from −1 to −0.4 V, the device evolves from strong to weak depletion and the upward band bending is gradually reduced [**Fig. 8(c)**]. Because the mobile-carrier population remains limited, the conductance loss is mainly attributed to carrier capture and emission involving interface-related states, with a possible additional contribution from donor-like states whose occupancy is modulated under negative bias [**Fig. 8(d)**] [3]. The peak reduction may partly reflect a weaker donor-like-state contribution as depletion is relaxed. Near turn-on, from approximately −0.3 to −0.1 V, increasing channel carriers enhance the interface-trap response. Using a continuum model incorporating surface-potential fluctuations [12], an apparent interface-state density of $9–12\times10^{11}$ $eV^{-1}cm^{-2}$ is extracted, while the corresponding time constant remains nearly constant at approximately 250 ns [**Fig. 8(e)**].

From weak to moderate accumulation ($V_G$=0–0.7 V), the bands bend further downward and $E_C-E_F$ decreases, allowing an increasing carrier population to interact with shallow and interface states [**Figs. 9(a)–9(c)**]. At $V_G$=0 V, the high-frequency peak remains close to the depletion-regime interface-related response, while an additional broad low-frequency component emerges. With increasing $V_G$, the two responses merge into one peak that first increases and then decreases. This evolution is consistent with a dominant contribution from shallow states in the IGZTO bulk, superimposed on interface-related and residual RC responses. Assuming a uniform distribution throughout the 4.5-nm-thick channel, the response remaining after subtraction of the interface-state background was converted into an apparent volumetric DOS. The extracted DOS exhibits a Gaussian-like profile with values of $10^{18}–10^{19}$ $eV^{-1}cm^{-3}$ and response times ranging from microseconds to nanoseconds [**Fig. 9(d)**]. The approximately 0.3 eV surface-potential modulation is comparable to the derived $E_C-E_F$, supporting its assignment to shallow states below $E_C$. Although the extracted DOS still contains a residual RC contribution, the weak correlations between $f_{peak}$ and $f_{pred}$, and between the extracted time constant and $R_{total}$, indicate that RC effects are limited in this bias range [**Figs. 9(e) and 9(f)**]. Accordingly, a rigorous equivalent-circuit model needs to be developed to fully separate residual distributed RC effects from the trap-related responses.

From moderate to strong accumulation ($V_G$=0.8–2 V), interface-related states again dominate the conductance response [**Figs. 9(g)–9(i)**]. The extracted interface-state density initially decreases and subsequently increases to $2.8–3.1\times10^{12}$ $eV^{-1}cm^{-2}$, whereas the time constant remains near 250 ns [**Fig. 9(j)**]. The initial reduction may arise from coupling with neighboring high-frequency responses, while the subsequent increase may reflect more complete carrier exchange with interface-related states as the accumulated electron density increases. The resulting minimum identifies a bias point with reduced conductance loss within the investigated frequency window, although its origin requires further clarification, for example, by low-temperature conductance measurements [16]. The approximately threefold difference in the apparent interface-related state density between accumulation and depletion may help clarify the distinct trap-mediated degradation mechanisms under positive and negative bias in AOS transistors.

Finally, **Figs. 10(a)** and **10(b)** compare the bias–frequency-resolved $G_p/\omega$ maps of CSC-IGZTO transistors with L=5 μm and widths of 80 and 50 μm, respectively. The reduced effective channel area lowers the overall trap-induced conductance loss, while preserving a similar bias–frequency distribution. In depletion, the interface-related peak remains nearly unchanged in magnitude and frequency. A stronger and broader response emerges from weak to moderate accumulation, followed by the reappearance of an interface-dominated response under strong accumulation. After area normalization, both devices exhibit consistent Dit values and bias dependence, with response times remaining near 250 ns [**Fig. 10(c)**], supporting the reproducibility of the extraction.

## IV. CONCLUSION

We propose a CSC AOS MOS test structure that overcomes the limitations of conventional structures and enhances the sensing of distinct trap-induced conductance-loss components in AOS transistors. By accounting for leakage, resistance, and device-geometry effects, the proposed method resolves the bias-dependent contributions of interface-related and shallow states and enables extraction of their apparent densities and characteristic response times. The CSC structure provides a simple and scalable platform for defect characterization and for guiding the performance and reliability optimization of AOS transistors.

**Acknowledgment**: This work was supported by the National Natural Science Foundation of China (92564301), the Innovation and Development Fund of Hangzhou Chengxi Sci-Tech Innovation Corridor (134000-X02402/006) and the NUS-AISI Joint Research Initiative Fund. Kai Chen gratefully acknowledges financial support from the Qiushi Flying Eagle Program of Zhejiang University.

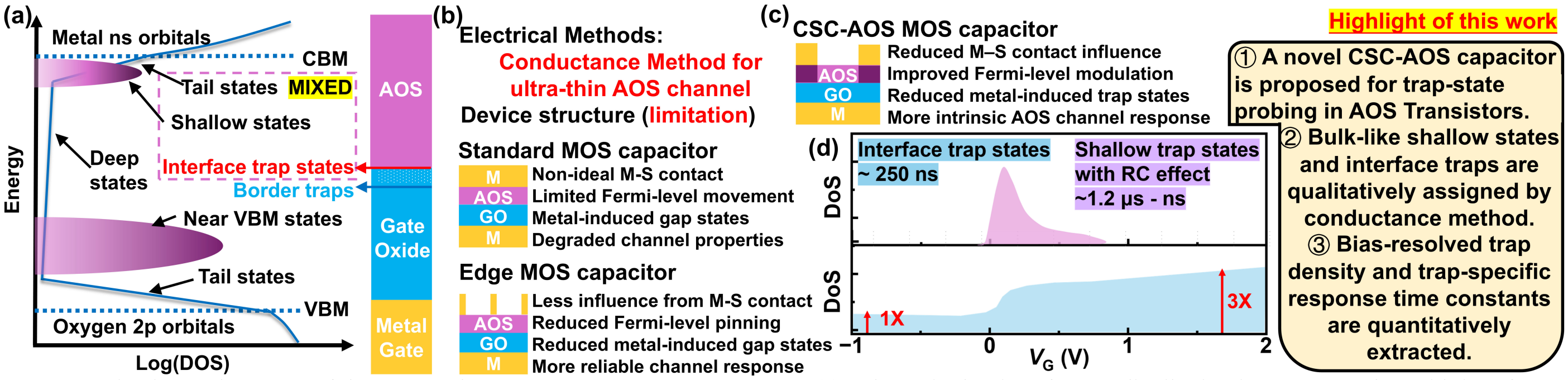

**Fig. 1**. Motivation and concept of the proposed CSC-AOS trap-state spectroscopy. (a) Schematic density-of-states distribution in AOS, together with interface and border traps in the gate-stack structure. (b) Structural limitations of conventional parallel-plate and edge MOS capacitors for conductance-based trap characterization in ultrathin AOS channels. (c) Proposed conductive–semiconducting–conductive AOS MOS capacitor, designed to reduce contact-related limitations and improve access to channel-associated conductance loss. (d) Bias-dependent density-of-states (DOS).

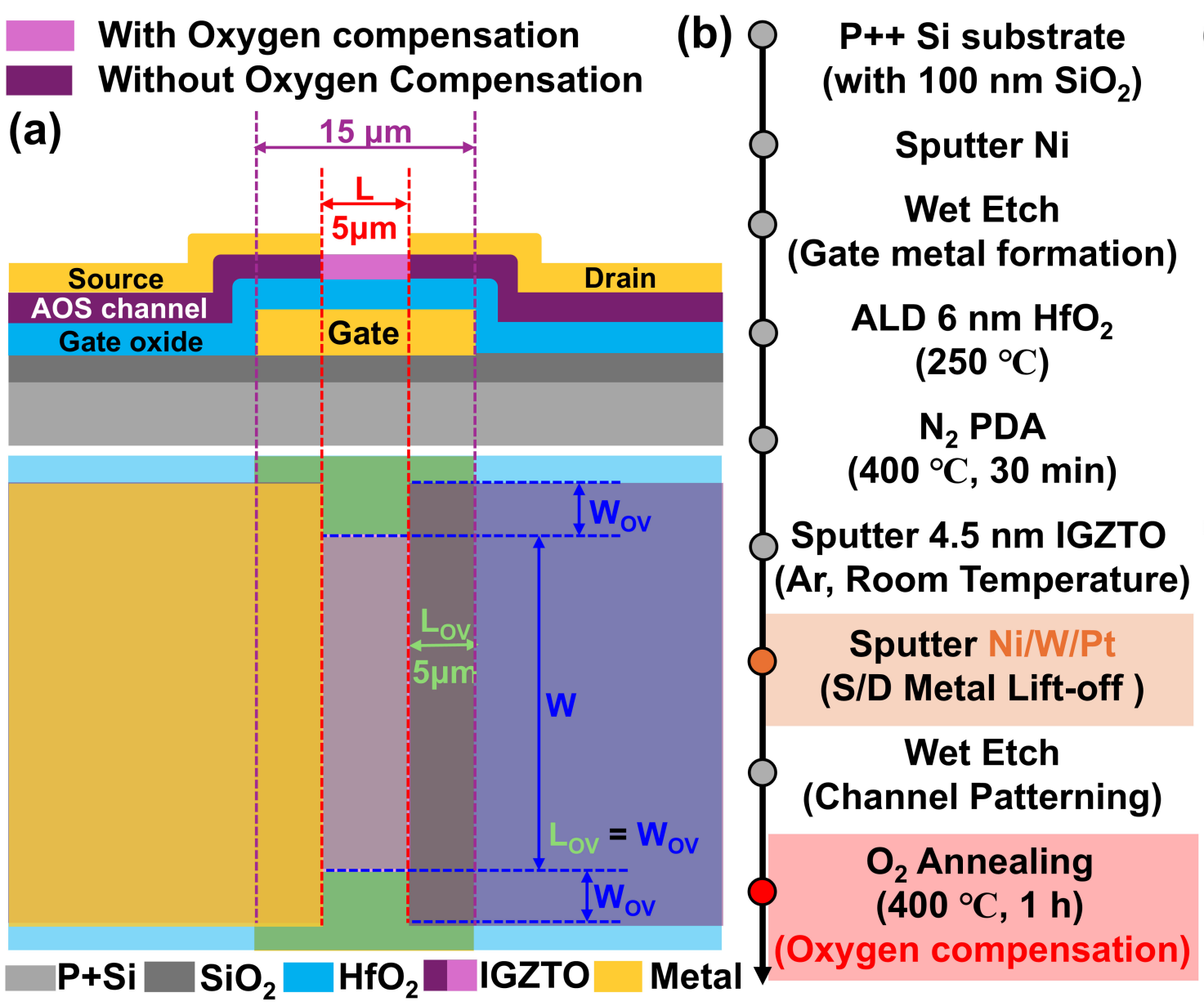

**Fig. 2**. (a) Cross-sectional and top-view schematics of the bottom-gate CSC-IGZTO transistor, showing the gate-to-source/drain overlap regions and detailed device dimensions. (b) Corresponding device fabrication process flow.

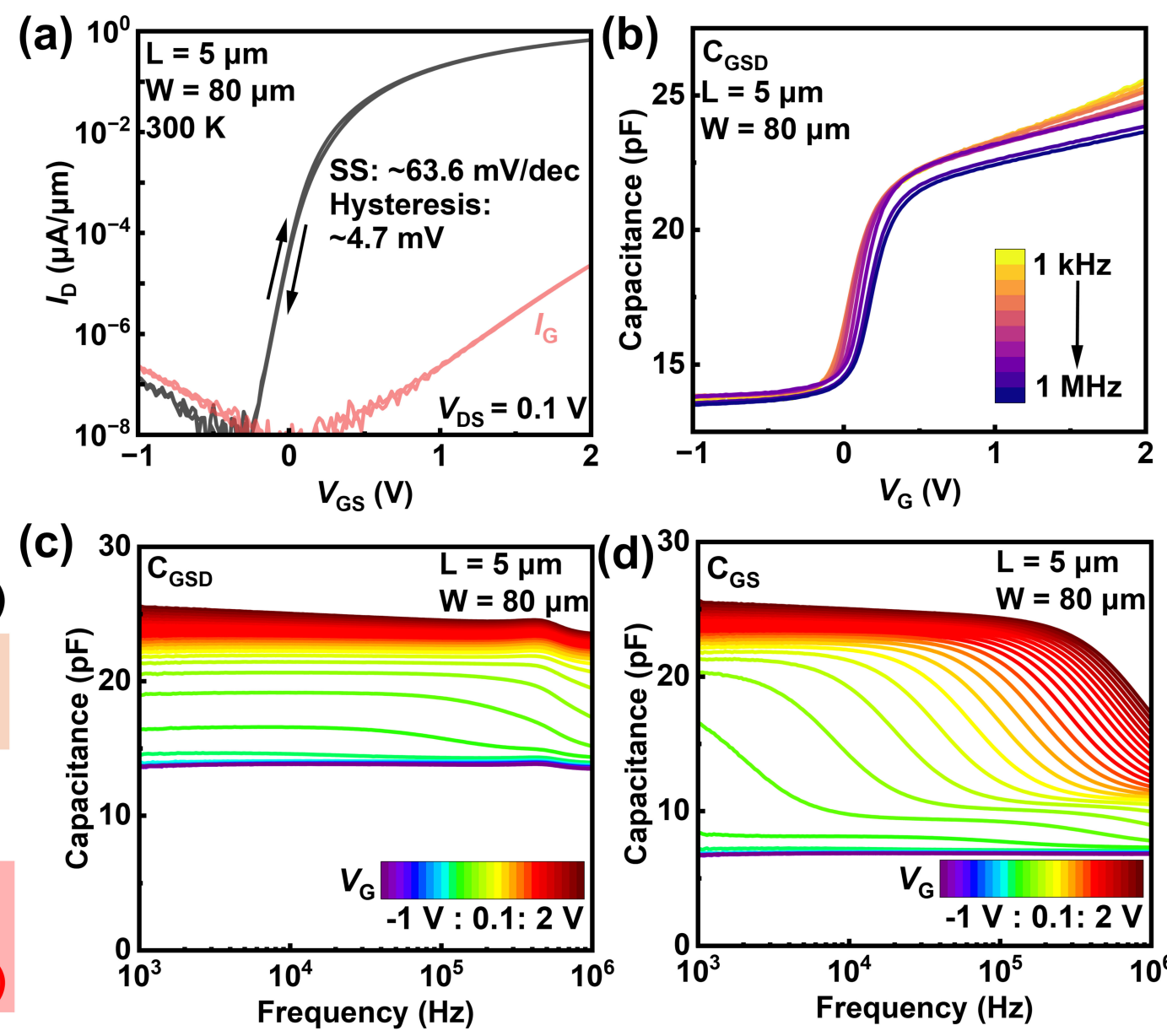

**Fig. 3**. (a) Measured transfer curve of the CSC-IGZTO transistor. (b) C–V characteristics with the S/D shorted. (c) Corresponding C–f characteristics. (d) C–f characteristics measured with the drain terminal floating.

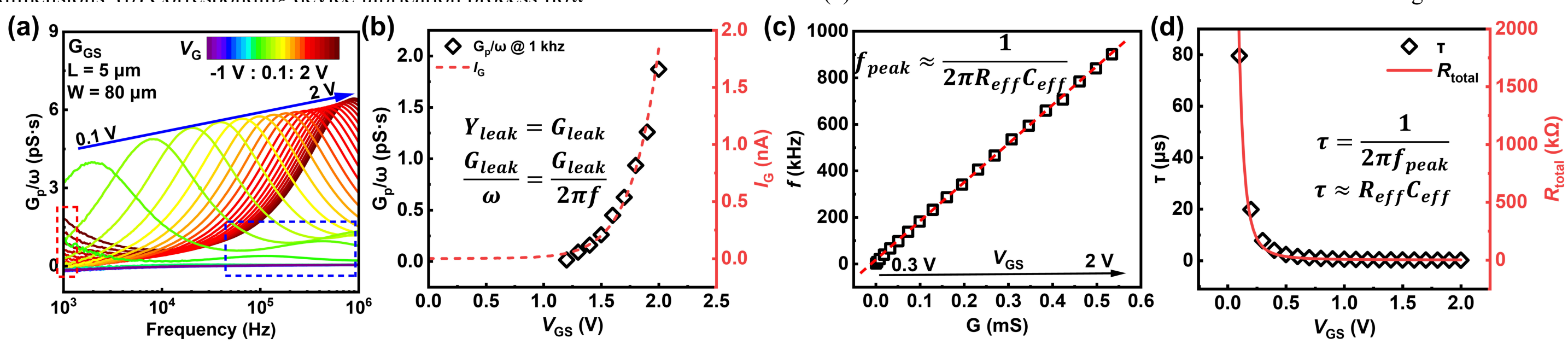

**Fig. 4**. (a) Gp/ω–f characteristics with the drain floating at different $V_G$. (b) Comparison of the 1 kHz $G_p/\omega$ component and the DC gate leakage versus $V_G$. (c) Extracted peak frequency as a function of total conductance ($1/R_{total}$). (d) Extracted time constant versus total resistance at different $V_G$.

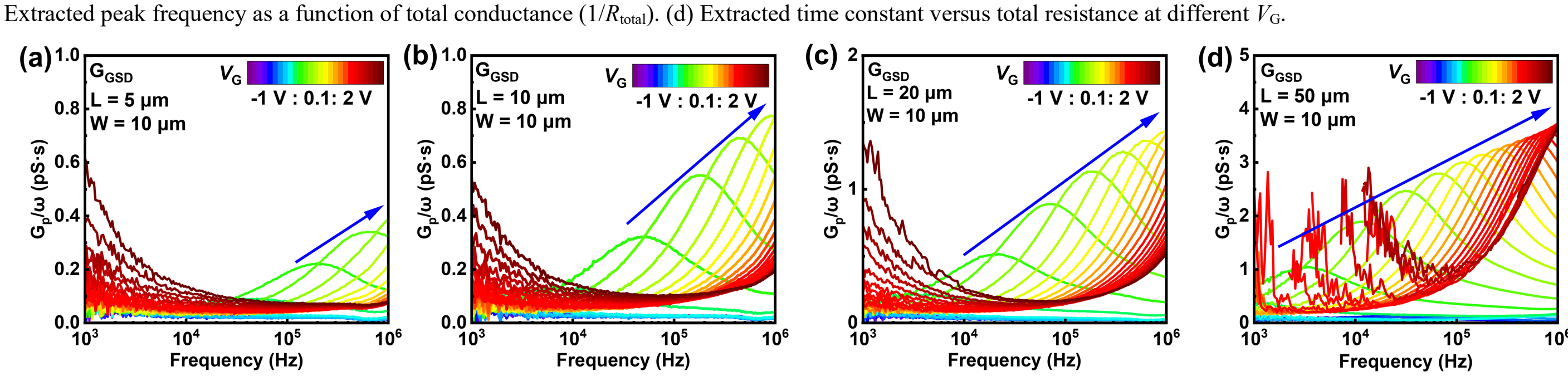

**Fig. 5**. (a-d) Gp/ω–f characteristics of CSC-IGZTO transistors with a fixed gate width of 10 μm and different gate lengths, measured from 1 kHz to 1 MHz.

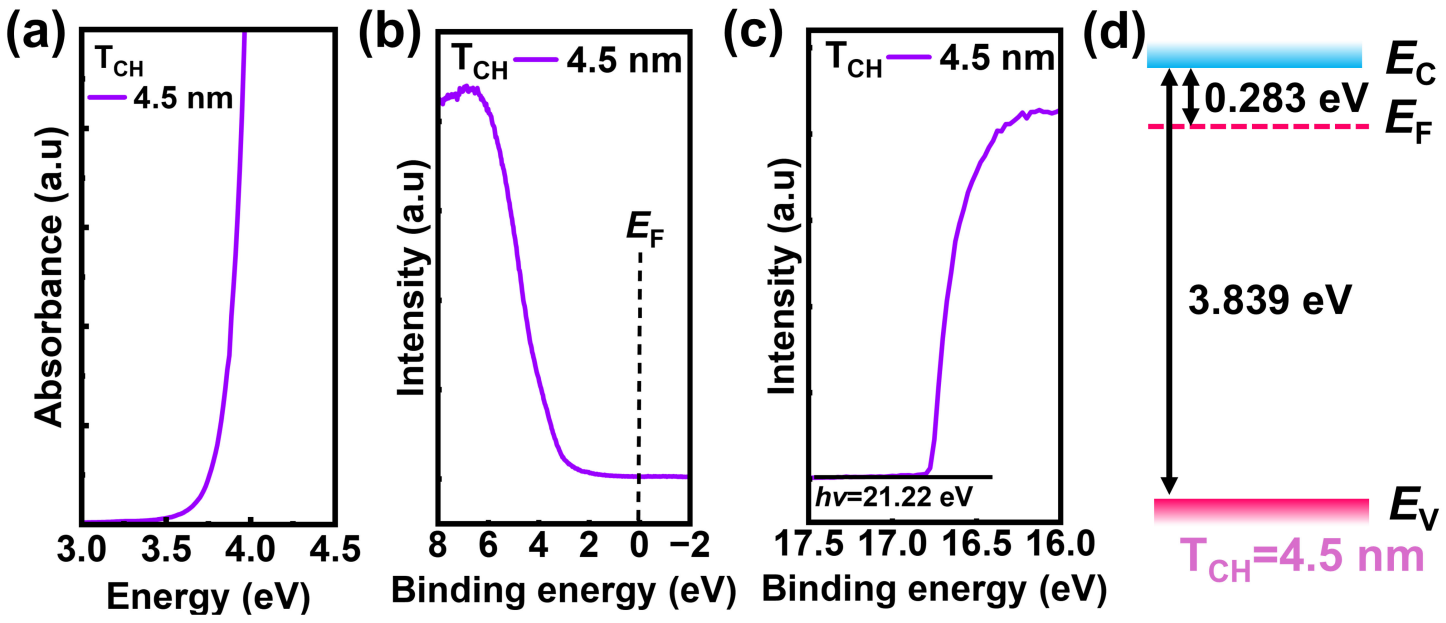

**Fig. 6**. (a) Optical absorption spectrum of the 4.5-nm-thick IGZTO film. UPS spectra near the Fermi level (b) and secondary-electron cutoff region (c). (d) Energy-band diagram extracted from the optical absorption and UPS results.

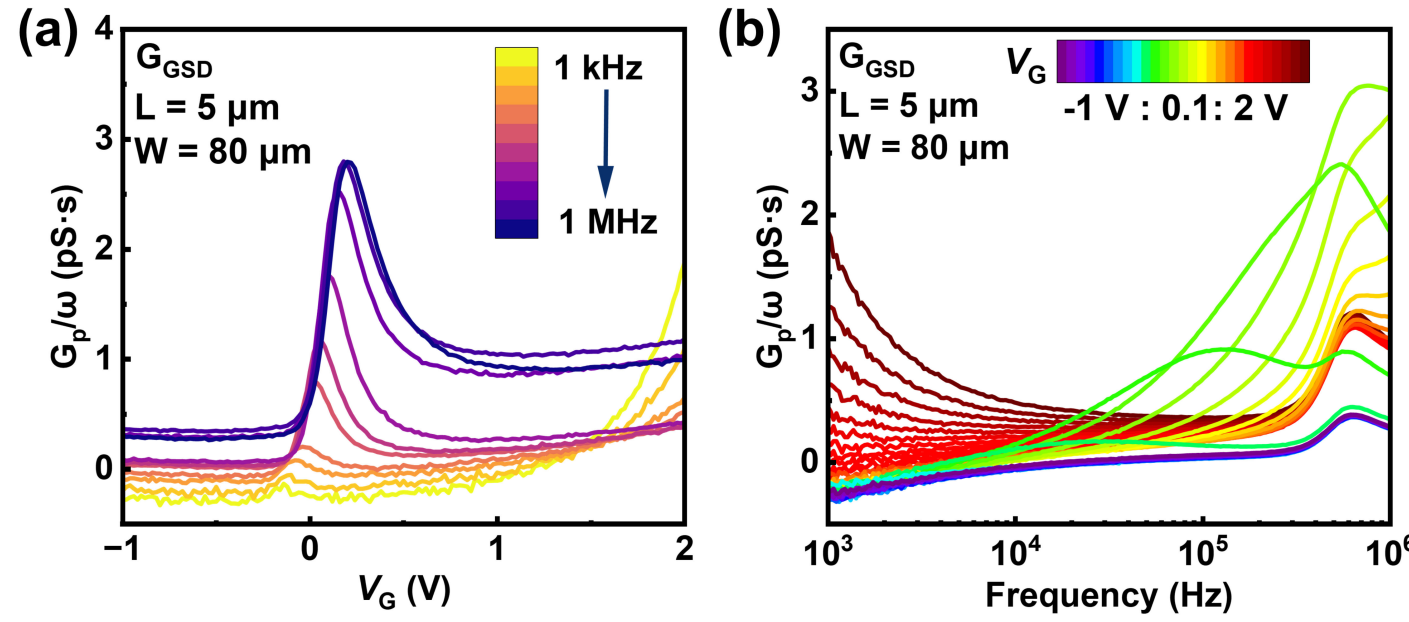

**Fig. 7**. (a) Gp/ω–$V_G$ characteristics of the CSC-IGZTO transistor with L=5 μm and W=80 μm, measured from 1 kHz to 1 MHz. (b) Corresponding Gp/ω–f characteristics measured from $V_G$=−1 to 2 V in 0.1 V steps.

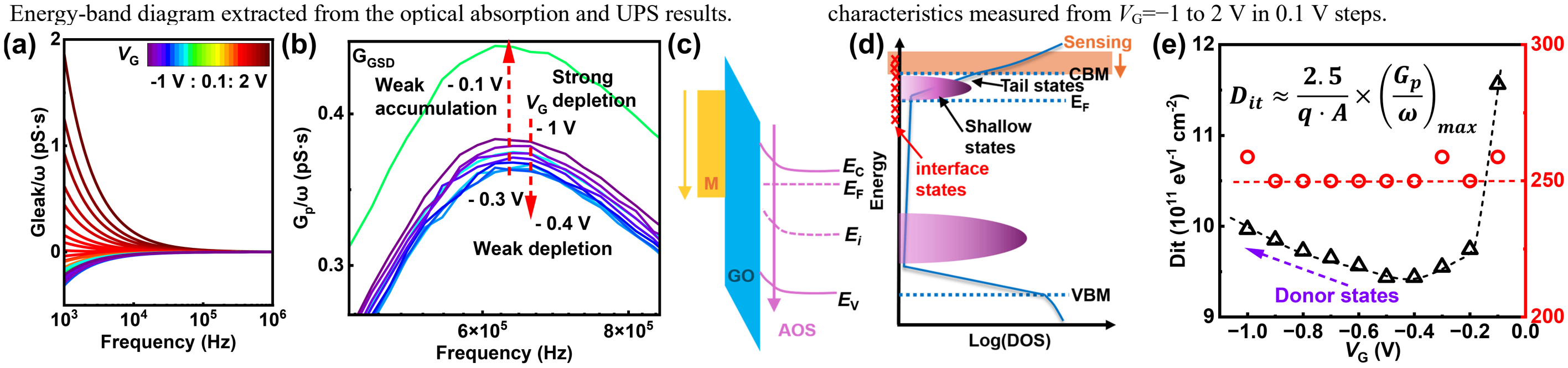

**Fig. 8**. (a) Estimated leakage contribution, $G_{leak}/\omega$–f, at different $V_G$. (b) Enlarged $G_p/\omega$–f spectra for $V_G$=−1 to −0.1 V. (c), (d) Schematics of the corresponding MOS band evolution and DOS-related conductance loss, respectively. (e) Extracted interface-state density and response time versus $V_G$.

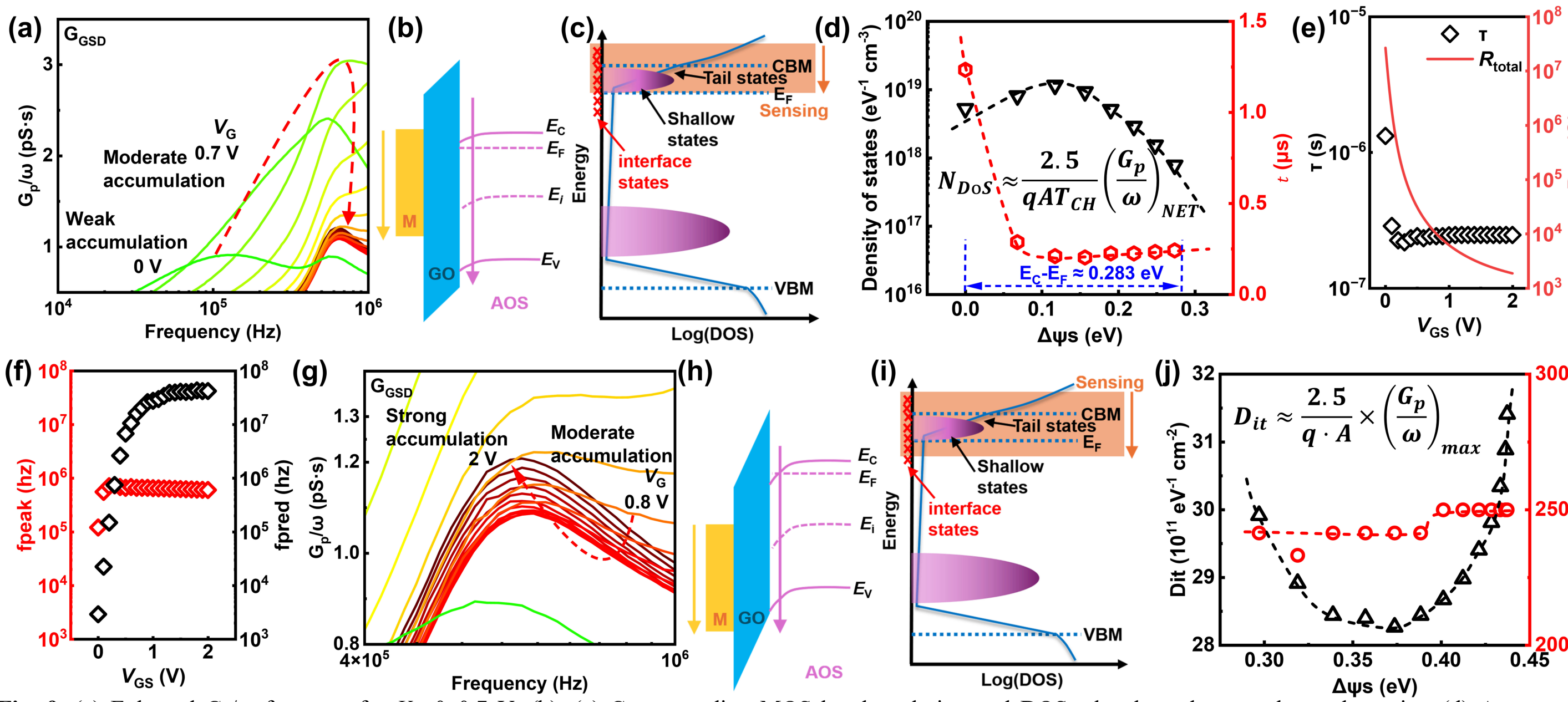

**Fig. 9**. (a) Enlarged $G_p/\omega$–f spectra for $V_G$=0–0.7 V. (b), (c) Corresponding MOS band evolution and DOS-related conductance-loss schematics. (d) Apparent shallow-state DOS and response time versus surface potential (ψs) after subtraction of the interface-state background, without RC correction. (e) Extracted time constant and total resistance versus $V_G$. (f) Comparison of fpeak and fpred. (g) Enlarged $G_p/\omega$–f spectra for $V_G$=0.8–2 V. (h), (i) Corresponding MOS band evolution and DOS-related conductance-loss schematics. (j) Extracted interface-state density and response time versus ψs.

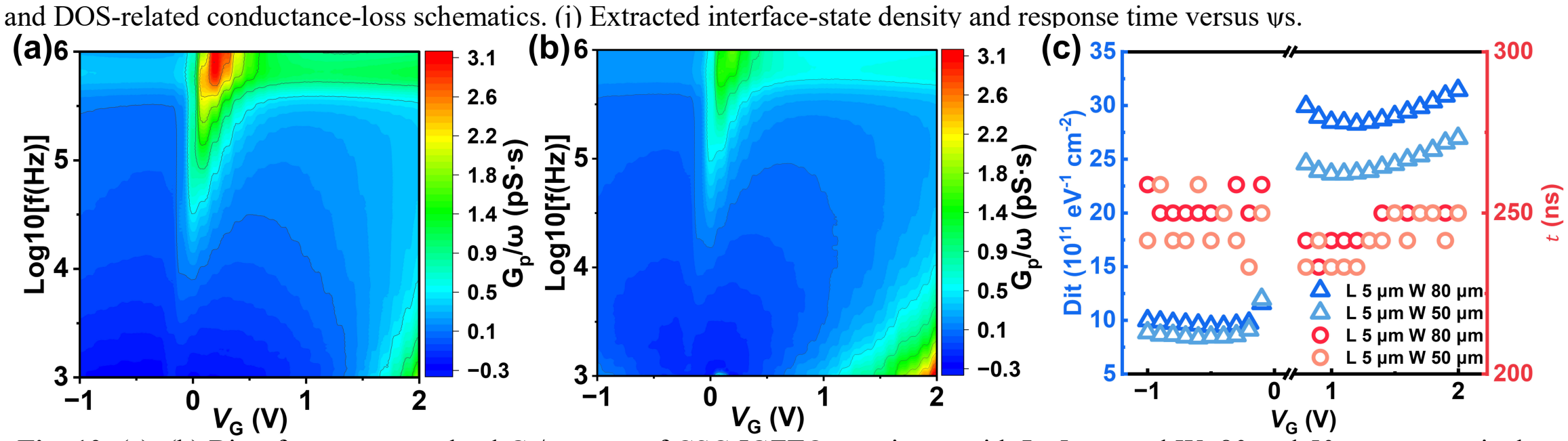

**Fig. 10.** (a), (b) Bias–frequency-resolved $G_p/\omega$ maps of CSC-IGZTO transistors with L=5 μm and W=80 and 50 μm, respectively. (c) Extracted interface-state density and response time versus $V_G$ for both devices, showing consistent bias dependence and comparable magnitudes.

References: [1] Nomura K et al. Nature, 2004. [2] Belmonte A et al. VLSI Symp. 2023. [3] Hosono Hideo et al. Wiley, 2022. [4] Belmonte A et al. Nat. Rev. Electr. Eng, 2025. [5] Liu S et al. IEDM, 2023. [6] Ye H et al., IEDM. 2020. [7] Chasin A et al. IEDM, 2021. [8] Zhao Y et al. IRPS, 2024. [9] Kamiya T et al. J. Display Technol, 2009. [10] Kamiya T et al., Phys. Status Solidi C, 2008. [11] Mattsson M J et al. arXiv, 2026. [12] Nicollian E H, Goetzberger A. Bell Syst. Tech. J, 1967. [13] Qu Y et al. TED, 2020. [14] Gaur A et al. 2D Materials, 2020. [15] Tang H et al. Solid-State Electron, 2024. [16] Martens K, et al. TED, 2008.